\documentclass[11pt, letter]{article}
\usepackage{times}
\usepackage{latexsym}
\usepackage{booktabs}
\usepackage{url}
\usepackage{listings}
\usepackage{color}
\usepackage{graphicx}
\usepackage{appendix}
\definecolor{dkgreen}{rgb}{0,0.6,0}
\definecolor{gray}{rgb}{0.5,0.5,0.5}
\definecolor{mauve}{rgb}{0.58,0,0.82}

\usepackage{times}                                       
\usepackage{amsmath}                                    
\usepackage{amssymb}                                     
\usepackage[export]{adjustbox}
\usepackage{footnote}

\makeatletter
\let\NAT@parse\undefined
\makeatother
\usepackage{hyperref}  
\usepackage{cleveref}

\usepackage{geometry}
\newgeometry{vmargin={20mm}, hmargin={17mm,20mm}}   

\title{Influence of Communication Among Shared Developers on the Productivity of Open Source Software Projects
}

\author{
  Sairamvinay Vijayaraghavan \\ \and
  Jason (Jinxiao) Song \\ \and
  Terry Guan \\\and
  Seongwoo Choi \\ \and
  Sutej Kulkarni
}

\date{}
\begin{document}
\maketitle
\begin{abstract}
    Many software developers rely on open source software for developing their applications and writing their source codes. Measuring an independent project’s overall productivity is still an open problem for many technology companies. In this project, we address to bridge the gap of analyzing which are the most important features for prediction of a We have chosen to collect data from GitHub via their application programming interfaces (API) and analyze the data we gathered to understand the relation between the average time to close an issue and the features that we collected. Since most of the data we gathered were not Gaussian, we had to preprocess the data using outlier detection and applying transformations before statistical modeling. The best model we observed was polynomial regression with degree 5. Overall, we noticed that there are many aspects of software development that make developers increase their productivity.
\end{abstract}

\section{Introduction}
Open Source Software (OSS) has become ubiquitous and provides developers with a  free-to-use platform for open public collaboration. With project developers spread across the world, OSS projects bring in diverse perspectives which are important for the success and sustainability of a project. In terms of OSS, GitHub has become the most popular web-based open-source software for developers to store and manage their code as well as track and control changes to their code.\\

In software development, developers and users (in the form of outside external contributors) utilize GitHub as an issue tracker and code collaboration tool across various projects. It is critically important to understand that software development across various developers is more productive when there is communication amongst developers across different projects. In addition, the progress of software development can be demonstrated through communication within the team and even across outside teams not working on this project. This phenomenon can be observed when developers accurately respond to any code-related bugs (commonly termed an issue in  GitHub) report if there are any. \\

In this project, we intend to study the effect of increased communication via shared code development on the productivity of the projects on GitHub. We particularly intend to find the most significant factors that influence the productivity of a project. To analyze the productivity within a software development medium, we decided to leverage the metric of average latency of issue closure associated with a project (we consider a repository as one project). It is computed by calculating the sum of all the issue lifetime (time taken to close an issue) for every issue within a project and dividing it by the number of closed issues for a given repository.  For considering the communication across developers beyond the considered project, we measured a metric that measured cross-development across multiple open source repositories. This metric was the average (mean) number of repositories worked by the top ten contributors (based on their involvement in the current project) outside the current project. In addition to this, we also analyzed certain other standard metrics relating to a project such as the number of contributors (team size), age of a project (how long it is active since its creation), total closed issues, etc. The goal of this project is to investigate the correlation between shared development and productivity on GitHub and find the critical features that influence issue closure latency. \footnote{https://github.com/Sairamvinay/Github-Drilling-Shared-vs-Solo}

\subsection{Research Questions}
Research and practice of software development are performed under various assumptions, their relationship with hypotheses, and discovered facts for validation. Based on the extensive empirical studies we have investigated, we have constructed several hypotheses for this research:

\subsubsection*{Research Question I:} In general, we know that more people working on development can improve productivity. Hence, we want to discover whether in open source software development: In an OSS medium, we can observe in general that more people working on development tends to improve productivity (by solving the proposed problems faster), but will this trend persist whenever there are more people working on the same project. We seek to answer whether for any project: With a larger software development team (more developers working on the same project), is there effectively a lower issue closure latency in open source projects? 

\subsubsection*{Research Question II:} In terms of one project’s productivity, would the presence of more shared developers reduce the issue closure latency? For the research purpose of this project, we consider shared developers as those who have contributed to multiple projects other than the current working project during the current project’s age time frame. 

\subsubsection*{Research Question III:} We analyze whether the project’s average latency of issue closure time reduces when the number of developers communicates knowledge across projects in terms of sharing knowledge? We consider this since we firmly believe that working on multiple projects would facilitate the transfer of knowledge across developers and improve the productivity of a software development team.

\subsubsection*{Hypothesis I:} An increase in the number of developers working on the same project (team size) can lead to lower latency in resolving issues

\subsubsection*{Hypothesis II:} An increase in the number of shared developers indicates more communication and collaboration between them, which leads to lower latency in resolving issues

\subsubsection*{Hypothesis III:} The project’s average latency of issue closure time can be reduced due to having developers that communicate knowledge across projects in terms of sharing knowledge.

Based on the research questions, we attempt to validate our hypothesis on the role of communication during software development.

\section{Literature Review}
In this research project, our team has decided to investigate the role of communication among shared developers during open-source software (OSS) development and how different team sizes of software developers can affect the overall productivity of the software development. \\

In the beginning, our focus is to evaluate the OSS development process by modeling and evaluating issue closure latencies against various features pertaining to the project and cross-development of contributors. For example, in the paper \cite{krishnan2000empirical}, the author examines different factors that can positively affect quality and productivity, such as the personnel capability of developers, product size, software process factors, and usage of tools. The author specifies the quality of the product as a function of product size, personnel capability, usage of tools, process factors, and proportion of investment in front-end development. The quality measures the number of unique problems reported by customers that are being solved. The author uses nonlinear data analysis and Ordinary Least Squares for the estimation procedures which is the first model method that we intend to use. The paper provides several insights for managers of commercial software products. The results of the quality equation identify several drivers of product quality. For example, the authors find that size of the software product personnel capability of project team members significantly affects quality. \\

Furthermore, we want to explore the effectiveness of having the software developer that is involved in many development projects on the overall productivity of those projects. There is limited theoretical work that exists in this field pertaining to the OSS research community. From previous research \cite{vasilescu2016sky}, multitasking has proved highly beneficial to productivity across GitHub projects through load balancing, more efficient work practices, learning, and cross-fertilization. Through our project, we have chosen to investigate how multitasking by shared developers across projects can facilitate communication of knowledge resulting in a lower latency in issue closure but the paper provides inspiration to using shared developers as the feature. In paper \cite{box1964analysis}, the authors talk about the importance of multi-discussing where developers can contribute their comments on multiple issues and switch their discussions between various issues they are handling at any given point in time. The paper claims that the work efficiency of developers increases on a per-issue basis, as multiple issues are discussed simultaneously by these developers. However, they also argue that multi-discussing may involve developers re-allocating their time and focus, which can cause a hindrance to the efficiency involved in issue closure. The datasets they have used (obtained using GHTorrent) are relatively small and their study indicates that there is more research needed in this area. We also observe that \cite{jarczyk2018surgical} the researchers emphasized the influence of having multiple developers working on a project and its effect on the overall quality of software development. \\

Furthermore, we wants to see how does the communication factor affects productivity. The researchers illustrated that the integration of information technology could increase the developers’ productivity by communicating and collaborating much faster. Since communication among the developers is essential to increase productivity, it is strictly essential to many industries that care about software quality which inspires our curiosity on the communication factor. In \cite{jarczyk2018surgical}, the researchers have used GHTorrent for their data collection and utilized survival analysis to understand the factors that affect issue closure rates in OSS projects. However, the researchers did not consider how important communication was. Hence, although it provides a good overview of the research direction, we decided not to do survival analysis but instead use polynomial regression because there are many parameters to consider when determining the productivity of developers in OSS projects. \\

Eventually, we are firmly believed that team sizes has a positive correlation with productivity. For example, in the paper \cite{tohidi2006productivity}, has explored the effectiveness of team sizes in projects which address our research question 1. As many projects may depend primarily on team members and how much they contribute to the projects, the researchers emphasized how the organizational structure on team sizes may be affected by communication and how tasks can be improved and promote the efficiency of organizations. However, the researchers did not consider the performance metrics regarding how many issues were closed after successfully communicating with team members. They also did not consider how one developer with more experience in a higher hierarchy may affect the overall productivity of the other developers in the team. The researchers also asserted that a well-developed organization may still have defects in their products and delay communication among the developers. With the advent of DSCM’s (distributed source code management systems) like GitHub, the ease with which developers can switch between projects has dramatically increased. Developers can find themselves working on different projects due to dependencies between projects or projects being in similar domains, which becomes important to identify and evaluate the productivity of shared developers. By handling multiple projects, developers can be more efficient in resolving issues and can also pass on knowledge/communicate across projects they have been involved with. Specifically, we want to explore the effectiveness of having how the one software developer that is involved that involved in many development projects can influence the overall productivity of those projects.\\

In comparison to previous works, we can observe some research regarding cross-project developments \cite{vasilescu2016sky}\cite{jarczyk2018surgical} and product quality. However, those research did not propose analyzing latency of productivity among shared developers. Similarly, some works have analyzed the team size \cite{vasilescu2016sky}\cite{krishnan2000empirical} and other important metrics such as organizational structures \cite{tohidi2006productivity} to determine the productivity of developers. However, to the best of our knowledge, there is very little work done with analyzing the cross-development factor in GitHub and several other parameters such as team size and the number of issues, which are being used for predicting the issue closure latency. There have been separate works in predicting and analyzing the issue closure latency, but none of those works have investigated the cross-development factor. Similarly, communication parameters have been analyzed, but there is very minimal work using the same variable we defined for cross-development. In addition, most of the works did not consider many analysis methodologies that we considered when conducting this research; for instance, we could not observe many papers that used polynomial regression for analyzing regression of productivity. \\

Some works have analyzed the team size \cite{vasilescu2016sky}\cite{krishnan2000empirical} and other important metrics such as organizational structures \cite{tohidi2006productivity} to determine the productivity of developers. However, to the best of our knowledge, there is very little work done with analyzing the cross-development factor in GitHub and several other parameters such as team size and the number of issues, which are being used for predicting the issue closure latency. There have been separate works in predicting and analyzing the issue closure latency, but none of those works have investigated the cross-development factor. Our work bridges the gap between predicting latency of issues using communication-based metrics such as cross-development factors as defined by our team in addition to the number of collaborators and other key metrics. \\

\section{Methodology}
For our project, we chose quantitative analysis since we were primarily analyzing data with numeric values and we wanted to derive conclusions from our data by statistical modeling techniques. In comparison to our progress report, we improved the number of samples by collecting more data by crawling the GitHub-based APIs. For the inputs, we decided to use the project size, the team size (number of contributors), total closed issues, the average number of cross-development repositories by a contributor for each repository, the average comments, project age, and the number of forks as the independent variables(features). We plan to fit a simple regression model that validates our hypothesis based on the learned regression coefficients and the significance values tests for identifying the most critical features for predicting the latency of issues. We use statistical tests learned from the course, particularly the P-value significance for these analyses. We report the standard metrics used for regression models, such as the R squared metric. For solving the regression models, we only used the Ordinary Least Squares (OLS) method since it was the most widely used regression solution method. We leveraged more tools such as the inverse of an error function (erfinv) for transforming some of our input variables into a Gaussian Distribution. We will also explain some of the challenges that we faced and our choices for feature transforming techniques. \\
 
Our design for modeling was straightforward just like how regular statistical models are trained. We collect our data from Github API, pre-process using outlier detection techniques and then transform some features for changing their raw distribution into a Gaussian-like distribution. Finally, we scaled our data using centering techniques and then predicted using a polynomial regression model.

\subsection{Data}
\subsubsection{Data Extraction: Features and Response Variables}
We have decided to investigate public repositories that have been created during the last five years (2016 - 2021) from GitHub. The GitHub Search API provides the ability to gather certain information about each repository, such as repository creation time, the number of issues within a repository, whether the issues are closed or open, the creation date, the issue closure date (if closed), the project size, and the number of comments per issue. Since the GitHub API imposed a rate limit while extraction, each person within our team was delegated to gather one year’s worth of public repositories in terms of their creation date. To make sure we don’t miss any crucial information, we crawl the API for the creation date of a project each day of the year, on a monthly basis. Results from this search are returned sorted based on “best matches/most relevant” as defined by the API. GitHub describes their “best matches/most relevant” as “Multiple factors which are combined to boost the most relevant to the top of the result list.”\cite{githubdocs}\\
 
We then used the name of Repos and users in conjunction with GitHub issues Search API to gather the issue closure information, the contributors working on a repository, the commit logs associated with the repository, and each issue's open date and closure date. Each of these unique repositories was considered as individual projects for this project’s scope. With this data, we can compute the average time to close an issue easily by averaging the lifetime of all closed issues of a repository. In order to determine the cross-development factor, we leverage a metric of measuring the average number of projects other than the current project being worked on by each of the top ten contributors during the time frame of this current project. The project’s top 10 contributors (based on their increased involvement in this project) were found and then the number of repositories that have been created after this project in a chronological sense by each of these ten contributors were accumulated and then finally averaged across all the ten contributors. \\
 
We gathered GitHub’s top 50 projects each day as returned by the search API, which allowed us to collect 12,800 samples initially. We filtered our search then to identify those repositories with at least one closed issue and obtained around 10381 samples before processing.

\subsubsection{Data Preprocessing and Analysis}
There are many outliers in our data and all our features did not possess a perfect Gaussian distribution but the average number of comments. For the outlier detection, we performed a simple outlier removal technique entirely based on the percentile values of the features. We calculated the 97.5th percentile of each of the features we collected and removed values larger than the 97.5th percentile by using these as a cutoff to remove the outlier samples. We also calculated the 2.5th percentile of our entire dataset which was leveraged as a cutoff for removing those samples which have feature values that are lesser than its 2.5th percentile. After outlier detection, we had a final dataset of 6757 samples. \\
 
For the purpose of analyzing parameters and displaying the analyzed results using Gaussian distribution, we intended to transform our features into a normal distribution. For our transformations, we leveraged the square root, logarithmic, and cubic root for each corresponding parameter, and we did not use more than one transformation for converting the features into normal Gaussian-like distributions. We did a logarithmic transformation for the project size (in KB) and the number of forks. Similarly, we applied a square root method for transforming the cross-development factor and a cube root method for transforming the closed issues. \\
 
For the age and team size variables of a project, we decided to choose a slightly different method from regular transformation methods. We chose to apply the inverse of an error function (erfinv) which was widely believed to be particularly useful in statistical applications for converting uniform random numbers into Normal random numbers. For the age variable (measured in days), we scaled into a [0,1] range and then applied an erfinv transformation to obtain the Gaussian Distribution. Similarly, we applied the same technique for the team size feature which however did not exactly transform into a Gaussian distribution because of its more discrete set of values. \\
 
Finally, after all our transformations, we chose to apply a simple centering technique that ensured that our variable distributions had a mean of 0 and standard deviation close to 1. It was performed by subtracting the mean of the variable from the variable values and then dividing by the standard deviation. From the frequency distribution in histograms, there were six parameters (besides team size) out of seven parameters that showed a widely spread out Gaussian distribution, which was a promising result for prediction in polynomial regression indicating that the transformation applied were powerful in obtaining Gaussian distributions. Refer to Figure [\ref{fig:histo}].\\

\begin{figure}[!ht]
    \centering
    \includegraphics[width=\linewidth]{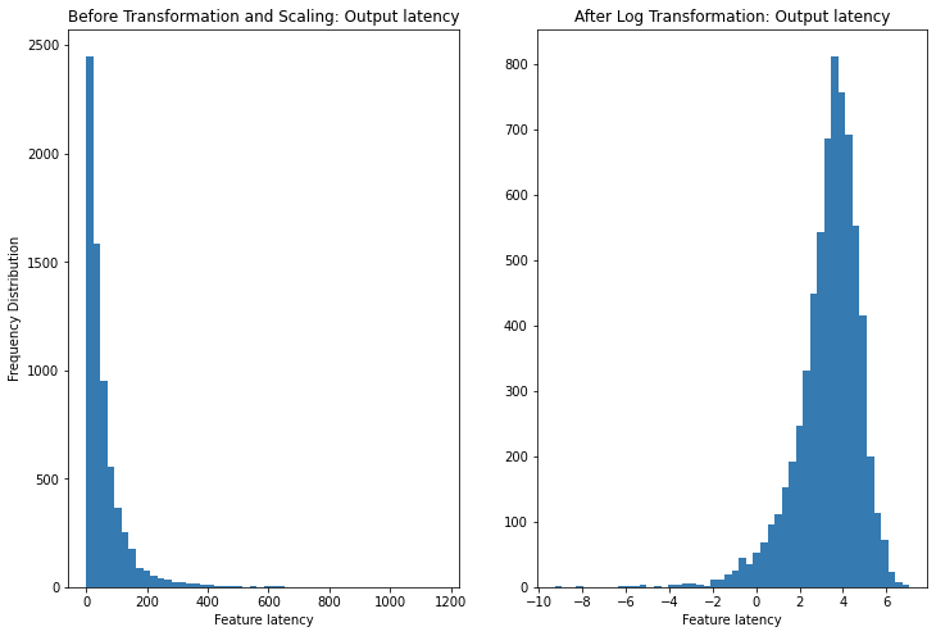}
    \caption{Output latency - Before and after Log Transformation}
    \label{fig:Latency}
\end{figure}

In addition to the features being transformed, we chose to model the latency after ensuring it possessed a normal Gaussian distribution, as referred to Figure [\ref{fig:Latency}]. We simply applied a log transformation on the average transformation and that was our output variable for predicting.

The graph in the appendix Figure [\ref{fig:scatter}] shows a scatter matrix among all variables collected. Although we can observe specific quick trends, more data need to be collected to verify this in a very general way. We can observe that with respect to the average latency, we can observe a slight downward trend with respect to the average number of repositories outside the current project and similarly another slight downward trend with respect to the number of contributors. Another observation was that our data displayed that there was no a visible linear relationship between latency and any of the input features and therefore we opted to work on a polynomial regression algorithm.

\subsection*{Methodologies Used:}
\subsubsection*{Ordinary Least Squares}
The ordinary least squares method is commonly used in statistics to determine the unknown parameters of a regression model. It calculates the sum of the squares of the residuals from the sample data-point values indicated by the polynomial function. However, as we worked on the OLS method, we observed that from our progress reports, transforming variables into a Gaussian distribution (which we incorporated recently) improved the modeling results. Having Gaussian distribution allowed us to analyze the dataset in more detail and it would be better for us to predict the latency of the developers, but we could not see any distribution and could not proceed with the analysis of the data. We did not consider adopting OLS for this research even though the OLS would be great for linear regression for observing linear relationships between input and output variables. 

\subsubsection*{Polynomial Regression}
A form of statistical analysis that uses the relationship between the dependent variable y and the independent variable x by fitting a polynomial function is called polynomial regression. It is a simple method that considers a simple polynomial function for modeling a response variable by solving the unknown parameters (simply coefficients) of the variables via optimization techniques. \\
 
We chose this method since it had a perfect balance between a simplistic method and non-linear behavior across the input and output variables. We chose this method over more complex methods such as Poisson regression and negative binomial regression since our data had a small number of samples and we still wanted to capture the non-linear relationships across the input and output variables. Hence we chose to proceed to model with these models for the scope of this project. We modeled our data via polynomial regression to explore the non-linear behavior of the features with respect to the response variable. Another advantage of using these methods allowed us to obtain more variables (although they were powered expressions of the features) for modeling. Hence we moved on to leverage polynomial regression to consider many dependent variables that may result in higher accuracy of predicting the response variable latency. Most of these analyses were conducted in Python (which includes crawling the API, pre-processing, and modeling our data).\\

\begin{figure}[!ht]
    \centering
    \includegraphics[scale=0.7]{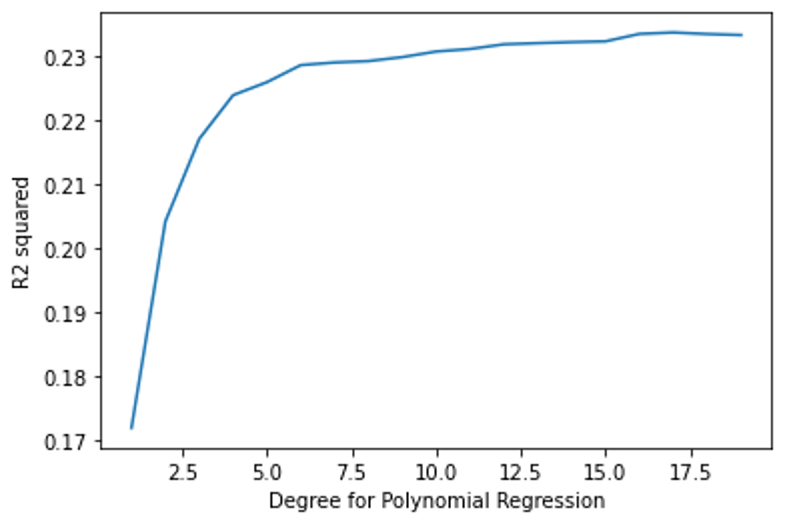}
    \caption{$R^2$-values}
    \label{fig:Rsquare}
\end{figure}

We noticed that from the scatter matrix Figure [\ref{fig:scatter}] most of the features bear a nonlinear relationship with the latency of closing issues, so we modeled using polynomial regression. We picked the degree of polynomial regression based on the performance of regression models of various degrees from 1 to 20 [\ref{fig:Rsquare}].

As we observed from the plot, we found that the polynomial regression models do not improve after a degree of 5 and hence we chose to stick with a small degree which also adds more non-linear behavior to our model. 

\section{Results}
\begin{figure}[!ht]
    \centering
    \includegraphics[width=1\linewidth]{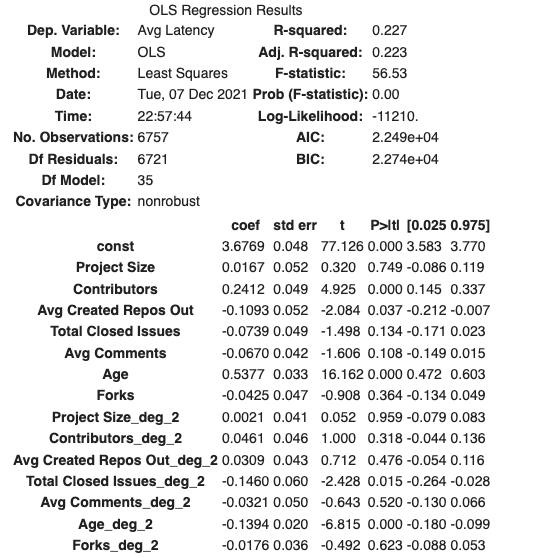}
    \caption{OLS Polynomial Regression results}
    \label{fig:OLS}
\end{figure}

We present the regression summary results we have obtained after fitting a polynomial regression model of degree 5 on the data features, see Figure [\ref{fig:OLS}].

The results show that based on the p-value significance (variables that have a significance $<$ 0.05))  indicates that the age of projects, contributors, cross-development factor (the number of repositories created outside a current project during its timeline), and total closed issues of the second degree are critical predictors for the response variable (log of the average latency).
 
We also find that for all the power raised for the features: forks, average comments, and forks are not very significant. The results confirm our hypotheses and in fact, we are able to validate our questions by confirming with the p-values of the features. Clearly, we can observe that the cross-development factor has a negative coefficient with respect to the output variable and hence it shows that an increase in the cross-development would decrease the average latency of closing issues. Similarly, we hypothesized that team size is correctly a strong predictor for latency. The positive coefficient for team size is a deduction where the results disagree with our hypotheses that it would bear a negative correlation with the output variable (that a larger team size would reduce the latency of closing issues).

\subsubsection*{RQ1:} Our results do not agree with our hypotheses of verifying whether a larger number of people within a team working on the same project during its lifetime would reduce the latency output variable. We strongly attribute this with the positive coefficient but, the variable of team size is definitely a strong predictor based on the p-value significance. We find that from our results that as a team size grows, we find an increase in the latency, which means it takes a longer time to close issues. We attribute the disagreement in the trend with what we observe and we hypothesize because we probably did not consider any other correlating factor (like weighting it based on the experience of each developer in the team) along with the team size which might have improved or shown a different trend.

\subsubsection*{RQ2:} The presence of shared developers was not directly gauged in our project; however, we consider the involvement of both the variables: forks and the cross-development factor for analyzing the presence of shared developers. We mainly focus on these variables because they have a straight connection with shared development: forks represent the number of various versions working on the same project and cross-development represents the number of other different projects worked during the same time. The results reveal that the forks do not have a huge influence on the latency while the cross-development factor does bear a negative relationship. We came to a conclusion that shared developers working more on this project do influence the latency of closed issues on the basis of these two factors. The negative coefficients of both these variables suggest that they bear a negative relationship as we had proposed in our hypothesis. We conclude that our results agree with the RQ2 we proposed.

\subsubsection*{RQ3:} There is a healthy agreement between the cross-development factor and the output variable of latency. We can observe that the cross-development factor indeed was a critical feature as a predictor for predicting latency and its coefficient was negative (-0.1093) which implied that an increase in this factor reduces the latency. This was in direct agreement with what we hypothesized and this research question was validated by our research results, which corresponds to the fact that cross-development between developers outside the current project is very helpful in resolving issues in open source software development because of possible increased sharing of knowledge across developers.

\section{Discussion}
\subsection*{Main Findings from Our Results}
Our final project results demonstrate that we have found the best predictors of latency, which was considered as a measure of productivity. Amongst those best predictors, we deduce that the average number of created repositories outside the current project by each of the top ten contributors is a strong variable that negatively influences the latency. This leads to a conclusion that cross-development across developers beyond the current project leads to more sharing of knowledge that allows more productivity and hence the developers are able to solve the issue quicker than the regular cases without shared development. In conclusion, we have validated our overall aim that collaboration between developers beyond the current working project improves the latency and hence increases productivity. \\

\subsection*{Strengths and Limitations}
We confirm that our project was successful in validating all our research questions and we were fairly successful in correctly stating and validating the relations in the form of our research questions. We had collected and modeled a simple dataset with a small set of features using polynomial regression. However, our results prove to be in good agreement with our hypotheses. While we have been successful in completing research on this topic, we wanted to explore more complex models such as Poisson and Negative Binomial with a larger dataset. Poisson distribution is a statistical procedure that shows the probability of a given number (events) of events happening in a fixed time interval and can use the results for estimating events of varying sizes (area or distance). We could have considered adopting the Poisson Distribution methodology to predict the number of days taken to resolve issues as a discrete variable. Also, we didn't apply Box-Cox Transformation in our features although it was one of the transformation methods of transforming non-normal dependent variables into a normal distribution. This was because it employed a complicated methodology and we didn't want to explore a very complicated function into our simplistic dataset. Similarly, we did not consider certain critical features such as the experience of developers, lines of code, and a number of edited files for modeling. Also, we considered a plain domain and did not consider our work within a particular domain: such as pertaining to only application development or entirely based on back-end development, etc.\\

\section{Conclusion}
We have seen several aspects of interpreting the relationship between the productivity of the developers in the open-source software and the collaboration across various developers.,  From our quantitative analysis, based on the data we have collected from the open-source software, GitHub, there are a few parameters that influence the latency of closing issues within software development. We have used certain statistical methodologies to analyze the data we collected from the scripts we have written and we have shown how to model the data we collected for a better understanding of the dataset and for better prediction of the latency of software development among the shared developers. Most importantly we have utilized simple but effective analysis methodologies such as Polynomial regression of a small degree 5 to predict the latency and we have observed that our method has provided many unique results, and have shown how to predict and measure the quality of open-source software. For further research, we can observe more on polynomial regression results and measure the actual results in a specific time period of the development and may apply other statistical techniques for a better understanding of productivity. \\

\section{Team Membership and Attestation of Work}
Seongwoo Choi, Terry Guan, Sutej Kulkarni, Jason (Jinxiao) Song, and Sairamvinay Vijayaraghavan agree to participate and work on this project with the understanding of the project.
\newpage
\bibliographystyle{plain}
\bibliography{bibliography.bib}

\begin{thebibliography}{1}

\bibitem{githubdocs}
Searching issues and pull requests.
\newblock {\em GitHub Docs}.

\bibitem{box1964analysis}
George~EP Box and David~R Cox.
\newblock An analysis of transformations.
\newblock {\em Journal of the Royal Statistical Society: Series B
  (Methodological)}, 26(2):211--243, 1964.

\bibitem{jarczyk2018surgical}
Oskar Jarczyk, Szymon Jaroszewicz, Adam Wierzbicki, Kamil Pawlak, and Michal
  Jankowski-Lorek.
\newblock Surgical teams on github: Modeling performance of github project
  development processes.
\newblock {\em Information and Software Technology}, 100:32--46, 2018.

\bibitem{krishnan2000empirical}
Mayuram~S Krishnan, Charlie~H Kriebel, Sunder Kekre, and Tridas Mukhopadhyay.
\newblock An empirical analysis of productivity and quality in software
  products.
\newblock {\em Management science}, 46(6):745--759, 2000.

\bibitem{tohidi2006productivity}
Hamid Tohidi and Mohammad~Jafar Tarokh.
\newblock Productivity outcomes of teamwork as an effect of information
  technology and team size.
\newblock {\em International Journal of Production Economics}, 103(2):610--615,
  2006.

\bibitem{vasilescu2016sky}
Bogdan Vasilescu, Kelly Blincoe, Qi~Xuan, Casey Casalnuovo, Daniela Damian,
  Premkumar Devanbu, and Vladimir Filkov.
\newblock The sky is not the limit: multitasking across github projects.
\newblock In {\em Proceedings of the 38th International Conference on Software
  Engineering}, pages 994--1005, 2016.

\end{thebibliography}
\newpage

\clearpage
\appendix
\appendixpage
\addappheadtotoc
\begin{figure}[h!]
    \centering
    \includegraphics[width=1\textwidth]{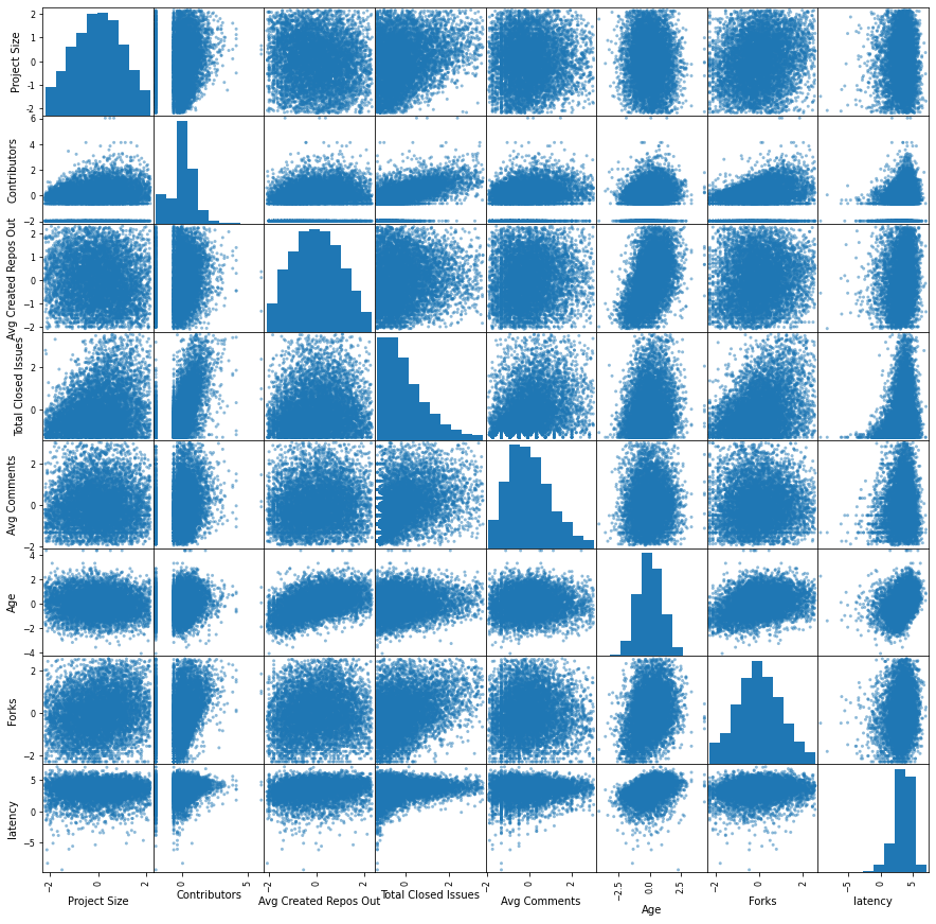}
    \caption{Scatter matrix among all the features and output}
    \label{fig:scatter}
\end{figure}

\newpage
\begin{figure*}[htp]
    \centering
    \includegraphics[width=1\textwidth]{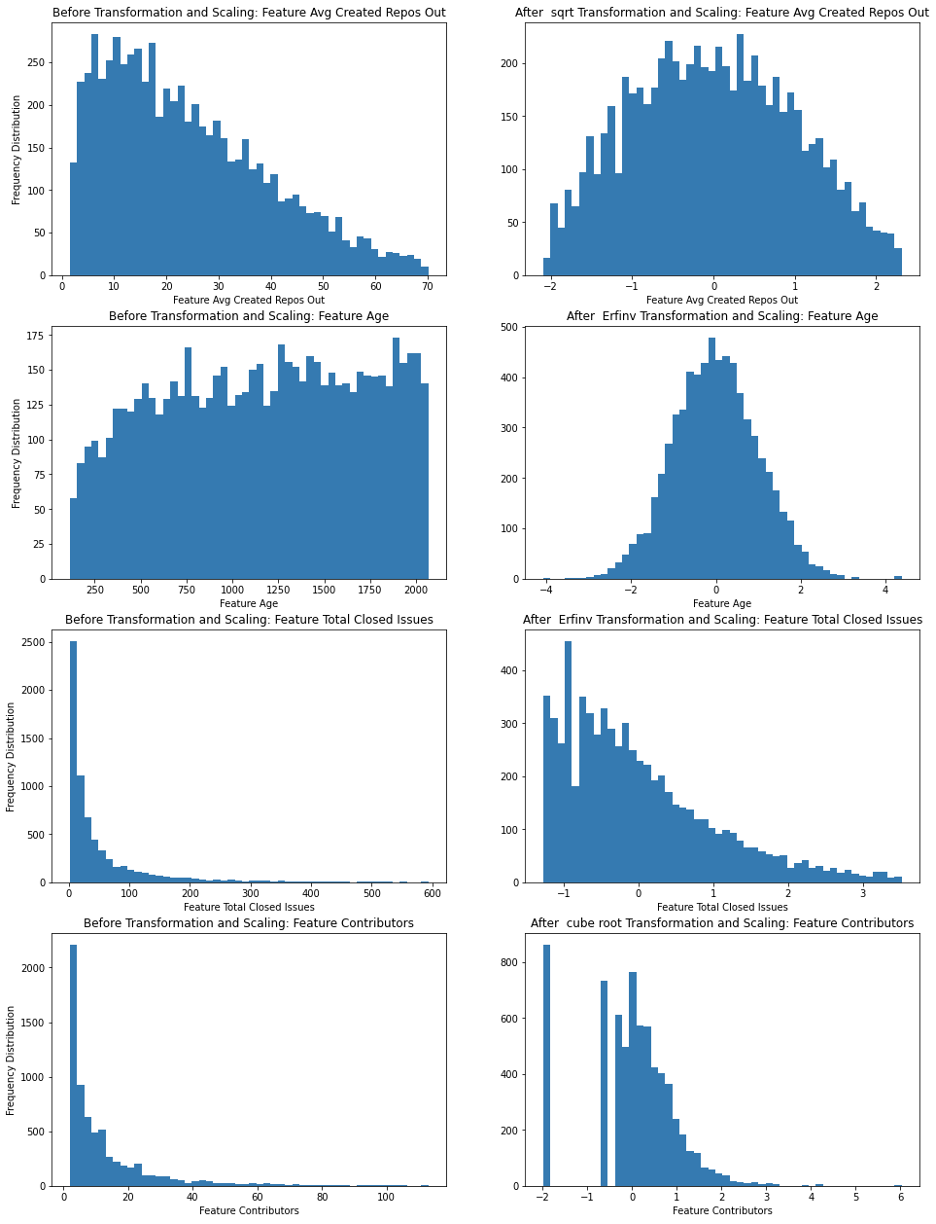}
    \caption{Frequency distribution in histograms}
    \label{fig:histo}
\end{figure*}

\clearpage

\end{document}